\documentclass[doublecol]{epl2}
\usepackage{amsmath}
\usepackage{amssymb}
\usepackage{esvect} 
\title{Rare events in networks with internal and external noise}
\shorttitle{Title} %Insert here a short version of the title if it exceeds 70 characters

\author{J. Hindes\inst{1} \and I. B. Schwartz\inst{1}}

\institute{                    
  \inst{1} U.S. Naval Research Laboratory, Code 6792, Plasma Physics Division - Washington, DC, U.S.\\
}
\pacs{64.60.aq}{Networks}
\pacs{05.45.-a}{Nonlinear dynamics}
\pacs{87.23.Cc}{Population dynamics}

\abstract{We study rare events in networks with both internal and external noise, and develop a general formalism for analyzing rare events that combines pair-quenched techniques and large-deviation theory. The probability distribution, shape, and time scale of rare events are considered in detail for extinction in the Susceptible-Infected-Susceptible model as an illustration. We find that when both types of noise are present, there is a cross-over region as the network size is increased, where the probability exponent for large deviations no longer increases linearly with the network size. We demonstrate that the form of the cross-over depends on whether the endemic state is localized near the epidemic threshold or not.}

\begin{document}
\maketitle
Many complex systems of interest can display rare and extreme events in their dynamics. Though such events occur infrequently, their existence may have drastic consequences for a system's long-term function and behavior. Examples include recurrent dynamics in excitable media with noise\cite{WeinanPNAS2006}, rogue waves in hydrodynamics and optics\cite{SolliNature2007,Bonatto2017PRE,Vanden2017Arxiv}, fixation in evolutionary games\cite{MobiliaEPL2010}, viral clearance\cite{Chaudhury2012}, and switching in quantum mechanical oscillators\cite{Dykman2015}. For systems with explicit noise dependence, large-deviation theory (LDT) has been developed, which can be used to predict the most likely pathway through which a rare event occurs\cite{FriedlinBook,AssafRev,SchwartzPRE2017}. The approach approximates well a broad class of rare dynamical processes described in the exponential tail of a probability distribution\cite{AssafRev,AghionPRL2017}. Some recent work has begun to apply this formalism to high-dimensional problems in network prediction and optimal control\cite{Motter2015PRX,Dykman1997PRE,HindesPRL2016,BianconiArxiv2017}. On the other hand, there is interest in understanding the role of network topology on rare events in discrete stochastic processes, where noise is implicit and internal, and the direct application of LDT techniques is less straightforward\cite{HindesPRE2017,MieghemPRE2015,HerrmannSR2017,HerrmannPRL2017}. In many situations, particularly in biology, a network may be subjected to both external environmental uncertainties and the inherent randomness of individual events\cite{KamenevPRL2008,DoeringPRE2017,LoraPRL2010}. Yet, no general methodology exists for analyzing rare processes in networks with internal and external noise sources -- the problem addressed in this letter. 

We first informally define a rare event as the appearance of some state in a network that is qualitatively different from its mean state, which occurs with exponentially small probability. Let us consider networks defined by a fixed set of (N) nodes and connections between them (called edges), such that they can be represented by adjacency matrices, $\bold{A}$, whose elements are binary, i.e., $A_{ij}\!=\!1$ if nodes $i$ and $j$ are connected and zero otherwise. In the following, we study a simple stochastic process explicitly, which models infection spreading over a contact network: the Susceptible-Infected-Susceptible model (SIS)\cite{PastorRMP2015}. However, such assumptions are illustrative only and can be generalized to other dynamical processes (as explained below), weighted networks, adaptive networks, etc. See for instance\cite{HindesPRE2017,HindesSR2017,HindesPRE2018}. 

In the SIS model nodes are divided into two types: infected and susceptible. Infected nodes transfer infection to their susceptible neighbors in a network at time, $t$, with probability-per-unit-time $\beta(t)$, and become susceptible with probability-per-unit-time $\alpha(t)$, where $\beta(t)$ and $\alpha(t)$ are known as the infection and recovery rates, respectively. In addition to the reactions on the network, $\beta(t)$ and $\alpha(t)$ may evolve according to independent stochastic processes. For illustration purposes, we take $\alpha(t)\!=\!\alpha\!=\!1$ and $\beta(t)\!=\!\beta_{0}[1-\xi(t)]$, where $\xi(t)$ is a zero-mean Gaussian white-noise process (GWNP) defined by the autocorrelation $\left<\xi(t)\xi(t')\right>\!=\!2D\delta(t-t')$, and $\beta_{0}$ is the average infection rate. Fluctuations in $\beta(t)$ model temporal variations in environmental conditions, close-proximity contact rates, etc.\cite{SalathePNAS2010,MendezPRE2012}.    

As long as $\beta_{0}$ is above threshold, $\beta_{0}\!>\!\beta_{c}$, an endemic state emerges in large networks with a metastable number of infected nodes, after an initial transient period of seed infection and growth. The threshold $\beta_{c}$ can be calculated approximately with a variety of mean-field techniques\cite{PastorRMP2015}; below, we use the so called ``pair quenched technique" (PQT), which is regarded as one of the most accurate for general network topologies\cite{MattaEPL2013}. 

\begin{figure}
\includegraphics[scale=0.236]{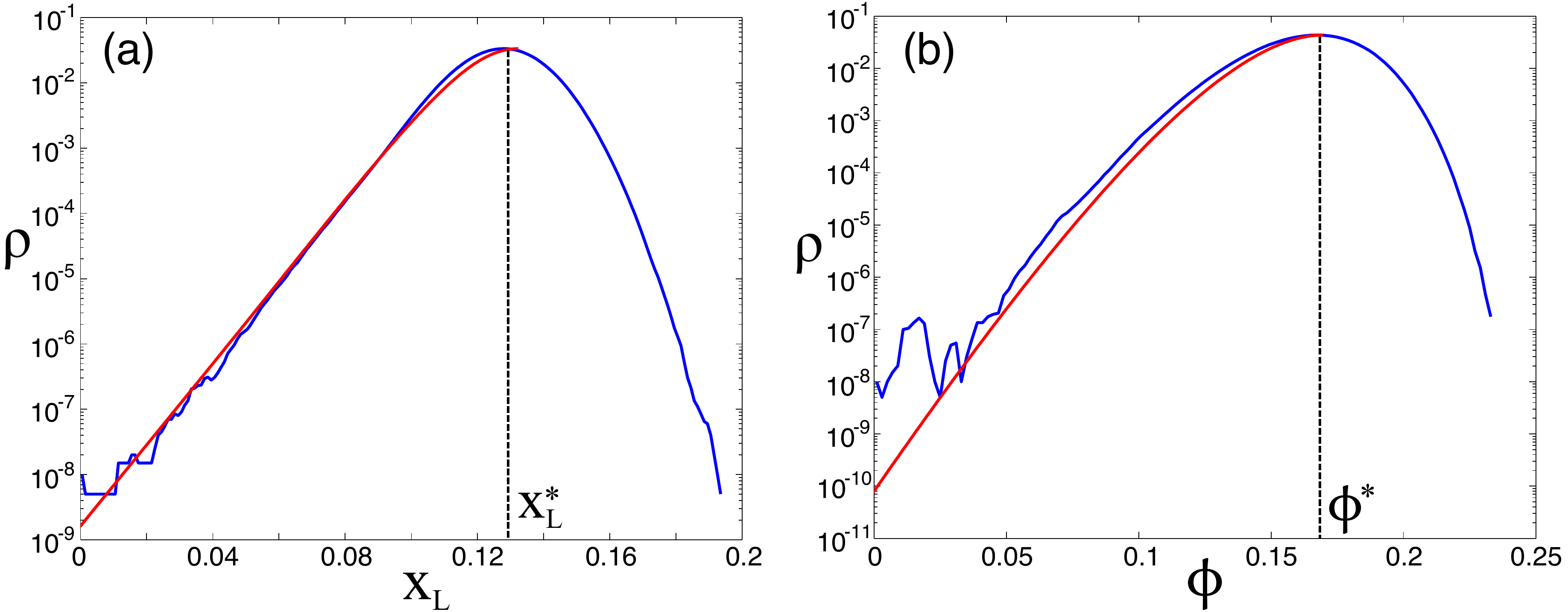}
\caption{Distributions of states in the SIS model with internal noise, $D\!=\!0$. (a) probability versus the fraction of infected nodes, $x_{L}$,  that are connected to the central node of a star network with $1001$ nodes and $\beta_{0}\!=\!0.160$. (b) probability versus the fraction of edges connecting infected and susceptible nodes, $\phi$, in a 6-regular network with $500$ nodes and $\beta_{0}\!=\!0.262$ \cite{FN2}. Endemic states are shown with a $(*)$. Predictions are shown in red.}  
\label{Distr}
\end{figure}

Though infection is metastable above threshold, a network continually fluctuates in time due to random perturbations in the infection rate and changes in nodal states, which we call {\it external} and {\it internal noise}, respectively. Over time, noise drives the network to explore rare states, the most drastic of which is {\it extinction} (in the SIS model), where the number of infected nodes goes to zero\cite{SchwartzJRSI2011,FerreiraPRE2016}. Preliminary insight on how such rare states emerge can be gained by examining histograms of time-series data from stochastic realizations of extinction\cite{HindesPRE2018}. Example histograms are shown in Fig.\ref{Distr} for two networks with $D\!=\!0$. As is typical for general networks, we find an {\it exponential probability distribution of large deviations} from the endemic state, which terminate at extinction. The distribution is peaked around the endemic state, and its shape depends on the infection rate and underlying network topology\cite{FerreiraPRE2016,HindesPRE2017}.

A theory of rare events on networks should allow for predictions of the observed distribution of network states leading to extinction and the expected rate of extinction. Our analytical approach is to derive an approximate time-evolution equation for the probability of network states in the SIS model that is consistent with PQT, and analyze its maximum-probability solutions. We emphasize that much of the formalism developed below can be applied to other rare processes, such as network switching\cite{HindesSR2017}. In fact, the primary difference among distinct classes of rare events resides in the boundary conditions needed to complete the theory and determine solutions of the LDT equations of motion (such as those derived below)\cite{SchwartzJRSI2011,SchwartzPRE2015}.   
%We work out our approach in $k$-regular and star networks explicitly, to illustrate rare events in weakly-connected homogeneous networks and heterogeneous networks showing localization, respectively. [Qualitative point summary].

As in \cite{HindesPRE2017,HindesSR2017}, we introduce a network ensemble, for analysis purposes, consisting of $C$ independent stochastic realizations of the SIS process on a fixed network. Ultimately, we are interested in the statistics of rare states in the limit of many realizations, or $C\!\rightarrow\!\infty$. Following PQT \cite{MattaEPL2013}, we describe the ensemble dynamics by the fraction of realizations in which node $i$ is infected at time $t$, $x_{i}(t)$, and the fraction of realizations in which node $i$ is susceptible and node $j$ is infected in an $i\!\!-\!\!j$ edge, $\phi_{ij}(t)$. These variables can be represented by an N-dimensional vector, $\vec{x}$, and an N$\times$N dimensional matrix $\boldsymbol{\phi}$.%Furthermore, we define auxiliary variables for the fraction of realizations in which both nodes $i$ and $j$ are infected, $\psi_{ij}(t)$, and susceptible, $\omega_{ij}(t)$. We note the useful constraints: $\psi_{ij}\!=\!\psi_{ji}\!=\!x_{j}-\phi_{ij}\!=\!x_{i}-\phi_{ji}$ and $\omega_{ij}\!=\!\omega_{ji}\!=\!1-x_{i}-\phi_{ij}\!=\!1-x_{j}-\phi_{ji}$, which can be used to eliminate redundant variables. For instance, if $x_{i}$ is specified for all $i$, then $\phi_{ij}$ needs to be specified only for distinct $i\!\!-\!\!j$ edges in the network\cite{MattaEPL2013}. 
       
The probability distribution for $\vec{x}$ and $\boldsymbol{\phi}$ in the ensemble at time $t$ given $\beta(t)$, $P(\vec{x},\boldsymbol{\phi},t;\!\beta)$, is governed by a master equation (ME). The ME is a probability-flux equation that depends on the rates of all possible reactions in the ensemble and the increments to $\vec{x}$ and $\boldsymbol{\phi}$ when a given reaction occurs. To see why the PQT is necessary, consider what happens when an infected node $j$ infects a node $i$. The fractions $x_{i}$ and $\phi_{ij}$ are incremented in a simple way: $x_{i}\!\rightarrow\!x_{i}\!+\!1/C$ and $\phi_{ij}\!\rightarrow\!\phi_{ij}\!-\!1/C$. However for additional neighbors of node $i$, e.g., node $l\!\neq\!j$, the increments to $\phi_{il}$ 
and $\phi_{li}$ depend on the state of $l$, which might be susceptible or infected. Assuming that we only know the states of nodes $i$ and $j$, then the state of node $l$ must be described by a probability. The probability that node $l$ is infected is given by the fraction of realizations in the ensemble in which the triplet $l\!-\!i\!-\!j$ has an infected node $l$, a susceptible node $i$, and an infected node $j$. To keep our description limited to one and two-node variables, $\vec{x}$ and $\boldsymbol{\phi}$, {\it we assume that the probabilities for states of nodes $l$ and $j$ in an $l\!-\!i\!-\!j$ triplet are independent}.   

The independence assumption enters when approximating the reaction increments as follows. When node $j$ infects node $i$
\begin{align}
\label{eq:Infection}
\boldsymbol{\phi}\overset{\!\!\!i-j}{\underset{\text{infect}}\longrightarrow}\boldsymbol{\phi}+\boldsymbol{\mathcal{I}}^{ij}\!/C.   
\end{align} 
We approximate the increment to $\phi_{il}$, $\mathcal{I}^{ij}_{il}$, by minus the probability that node $l$ is infected given node $i$ is susceptible (independent of $j$). Similarly, we approximate the increment $\mathcal{I}^{ij}_{li}$ by the probability that node $l$ is susceptible given node $i$ is susceptible (independent of $j$):       
\begin{align}
\label{eq:InfectionIncrements}
\mathcal{I}^{ij}_{il}&=-\phi_{il}A_{il}A_{ij}(1-\delta_{jl})/(1-x_{i}),\\
\mathcal{I}^{ij}_{li}&= (1\!-\!x_{i}\!-\!\phi_{il})A_{il}A_{ij}(1-\delta_{jl})/(1-x_{i}),\\
\mathcal{I}^{ij}_{ij}&=-A_{ij},\;\;\; \mathcal{I}^{ij}_{ji}=0. %\; \text{and}\;0\;\text{otherwise}.
\end{align}
On the other hand, if node $i$ recovers, $x_{i}\!\rightarrow\!x_{i}-1/C$ and  
\begin{align}
\label{eq:Recovery}
\boldsymbol{\phi}\!\overset{\!\!\!i}{\underset{\text{recover}}\longrightarrow}\!\boldsymbol{\phi}+\boldsymbol{\mathcal{R}}^{i}\!/C.  
\end{align} 
We approximate the increment to $\phi_{ij}$, $\mathcal{R}^{i}_{ij}$, by the probability that node $j$ is infected given node $i$ is infected, and approximate the increment $\mathcal{R}^{i}_{ji}$ by minus the probability that node $j$ is susceptible given node $i$ is infected:   
\begin{align}
\label{eq:RecoveryIncrements}
\mathcal{R}^{i}_{ij}&=(x_{i}\!-\!\phi_{ji})A_{ij}/x_{i}, \\
\mathcal{R}^{i}_{ji}&=-\phi_{ji}A_{ij}/x_{i}.
\end{align}
Altogether the increments give the ME for the network ensemble 
\begin{align}
\label{eq:MasterEquation}
&\!\!\frac{1}{C}\!\frac{\partial P}{\partial t}(\vec{x},\boldsymbol{\phi},t;\!\beta)=\alpha\!\sum_{i}\!\!\Big[x_{i}+\frac{1}{C}\Big]P\Big(\vec{x}\!+\!\frac{\vec{1}_{i}}{C},\boldsymbol{\phi}\!-\!\frac{\boldsymbol{\mathcal{R}}^{i}}{C},t;\!\beta\Big)\nonumber \\
&+\beta\!\sum_{ij}\!A_{ij}\Big[\phi_{ij}+\frac{1}{C}\Big]P\Big(\vec{x}\!-\!\frac{\vec{1}_{i}}{C},\boldsymbol{\phi}\!-\!\frac{\boldsymbol{\mathcal{I}}^{ij}}{C},t;\!\beta\Big)\nonumber \\ 
&-\!\alpha\!\sum_{i}\!x_{i}P(\vec{x},\boldsymbol{\phi},t;\!\beta)-\beta\!\sum_{ij}\!A_{ij}\phi_{ij}P(\vec{x},\boldsymbol{\phi},t;\!\beta),
%&\!\!\frac{1}{C}\!\frac{\partial P}{\partial t}(\vec{x},\boldsymbol{\phi},t;\!\beta)\!=\!-\beta\!\!\sum_{i,j}\!\phi_{ij}P(\vec{x},\boldsymbol{\phi},t;\!\beta)\!-\!\alpha\!\!\sum_{i}\!x_{i}P(\vec{x},\boldsymbol{\phi},t;\!\beta)\nonumber \\
%&\!\!+\beta\!\sum_{i,j}\!\!\Big[\phi_{ij}+\frac{1}{C}\Big]P\Big(\vec{x}\!-\!\frac{\vec{1}_{i}}{C},\boldsymbol{\phi}\!+\!\frac{\delta_{ij}}{C}\!+\!\!\sum_{l}\!\frac{\Delta_{ij,il}^{I}+\Delta_{ij,li}^{I}}{C},t;\!\beta\Big)\nonumber \\
%&\!\!+\alpha\!\sum_{i}\!\!\Big[x_{i}+\frac{1}{C}\Big]P\Big(\vec{x}\!+\!\frac{\vec{1}_{i}}{C},\boldsymbol{\phi}\!+\!\!\sum_{j}\!\frac{\Delta_{i,ij}^{R}+\Delta_{i,ji}^{R}}{C},t;\!\beta\Big),
\end{align}
where $\vec{x}+\vec{1}_{i}/C=(x_{1},x_{2},...,x_{i}+1/C,...,x_{N})$.

We would like to extract a solution of Eq.(\ref{eq:MasterEquation}) that describes rare events. The functional form for the probability of network states in individual stochastic realizations is suggested by Fig.\ref{Distr}. As remarked above such distributions, $\rho(\vec{x},\boldsymbol{\phi},t;\!\beta)$, are exponential functions of $\vec{x}$ and $\boldsymbol{\phi}$,  
\begin{align}
\label{eq:QuenchedDistribution}
\rho(\vec{x},\boldsymbol{\phi},t;\!\beta)=b\exp\{-S(\vec{x},\boldsymbol{\phi},t;\!\beta)\},
\end{align}
where $b$ is a normalization constant. As a consequence, the ensemble distribution takes a Wentzel-Kramers-Brillouin form (WKB), $P(\vec{x},\boldsymbol{\phi},t;\!\beta)\!=\!a\exp\{-CS(\vec{x},\boldsymbol{\phi},t;\!\beta)\}$: a product of independent and identical distributions. The WKB form is an essential property of the probability for rare processes describable by LDT\cite{AssafRev}. Substituting the WKB ansatz into Eq.(\ref{eq:MasterEquation}), Taylor expanding Eq.(\ref{eq:MasterEquation}) in powers of the small parameter $1/C$, and neglecting terms of $\mathcal{O}(1/C)$ or smaller \cite{HindesPRE2017,HindesSR2017,DykmanReview}, gives a Hamilton-Jacobi equation for the probability exponent $S(\vec{x},\boldsymbol{\phi},t;\!\beta)$:  
\begin{align}
\label{eq:HamiltonJacobi}
\frac{\partial S}{\partial t} + H\Big(\vec{x},\boldsymbol{\phi},\frac{\partial S}{\partial \vec{x}},\frac{\partial S}{\partial \boldsymbol{\phi}};\!\beta\Big)=0. 
\end{align}
As in analytical mechanics, $S$ and $H$ are known as the action and Hamiltonian, respectively. In particular, the Hamiltonian is a function of derivatives of $S$, defined as conjugate momenta 
\begin{align}
\label{eq:Momenta}
p_{i}=\frac{\partial S}{\partial x_{i}}, \;\;\text{and}\;\; m_{ij}=\frac{\partial S}{\partial\phi_{ij}}, 
\end{align}
with 
\begin{align}
\label{eq:Hamiltonian}
&H(\vec{x},\boldsymbol{\phi},\vec{p},\boldsymbol{m};\!\beta)=\alpha\!\sum_{i}\!x_{i}\bigg(\!\!\exp\!\Big\{\!\!-\!p_{i}\!+\!\sum_{j}\!\frac{A_{ij}}{x_{i}}\!\big[\!-\!\phi_{ji}m_{ji}\nonumber \\
&+(x_{i}\!-\!\phi_{ji})m_{ij}\big]\!\Big\}\!-\!1\!\bigg)+\beta\!\sum_{ij}\!A_{ij}\phi_{ij}\!\bigg(\!\!\exp\!\Big\{p_{i}\!-\!m_{ij}\nonumber\\
&+\!\sum_{l}\!A_{il}\frac{(1\!-\!\delta_{jl})}{(1\!-\!x_{i})}\!\big[\!-\!\phi_{il}m_{il}+(1\!-\!x_{i}\!-\!\phi_{il})m_{li}\big]\!\Big\}\!-\!1\!\bigg). 
\end{align}

It is important to note that the elements of $\boldsymbol{\phi}$ are not independent. This can be seen by considering the fraction of realizations in the ensemble in which both nodes sharing an edge are infected. For example, the fraction of such $i\!-\!j$ edges is $x_{j}-\phi_{ij}\!=\!x_{i}-\phi_{ji}$\cite{MattaEPL2013}, which implies $\phi_{ji}\!=\!x_{i}-x_{j}+\phi_{ij}.$ If such a constraint is enforced from the onset, it implies that $m_{ji}\!=\!0$ in Eq.(\ref{eq:Hamiltonian}), since the probability does not depend on $\phi_{ji}$ explicitly. We emphasize that when constructing a Hamiltonian from Eq.(\ref{eq:Hamiltonian}) using such constraints in $\boldsymbol{\phi}$, one must select a permutation for each $i\!-\!j$ edge in the network, e.g, $\phi_{ij}$, replace the other permutation $\phi_{ji}$ by $x_{i}-x_{j}+\phi_{ij}$, and set $m_{ji}\!=\!0$ in Eq.(\ref{eq:Hamiltonian}). This procedure is useful for eliminating redundancy. We demonstrate with examples shortly.
%As noted above, the edge variables are not independent, e.g., $\phi_{ji}\!=\!x_{i}-x_{j}+\phi_{ij}.$ If such a constraint is enforced from the onset, it implies that $m_{ji}\!=\!0$ in Eq.(\ref{eq:Hamiltonian}), since the probability should not depend on $\phi_{ji}$ explicitly. We emphasize that when constructing a Hamiltonian from Eq.(\ref{eq:Hamiltonian}) using constraints in $\boldsymbol{\phi}$, one must select a permutation for each $i\!-\!j$ edge in the network, e.g, $\phi_{ij}$, replace the other permutation $\phi_{ji}$ by $x_{i}-x_{j}+\phi_{ij}$, and set $m_{ji}\!=\!0$ in Eq.(\ref{eq:Hamiltonian}). We demonstrate with examples shortly.  

The solution of Eq.(\ref{eq:HamiltonJacobi}) for the probability exponent (action) is well known from analytical mechanics, 
\begin{align}
\label{eq:Action}
\!\!\!\!S(\vec{x},\boldsymbol{\phi},t;\!\beta)=\!\!\int\!\!\big[\vec{p}\cdot\dot{\vec{x}}\;+\!<\!\!\boldsymbol{m},\!\dot{\boldsymbol{\phi}}\!\!>\!-H(\vec{x},\boldsymbol{\phi},\vec{p},\boldsymbol{m};\!\beta)\big]dt,  
\end{align}
with scalar products $\vec{p}\cdot\dot{\vec{x}}\!=\!\sum_{i}p_{i}\dot{x}_{i}$ and $<\!\!\boldsymbol{m},\!\dot{\boldsymbol{\phi}}\!\!>=\!\sum_{ij}\!m_{ij}\dot{\phi}_{ij}$, and where $\vec{x}$, $\vec{p}$, $\boldsymbol{\phi}$, and $\boldsymbol{m}$ satisfy Hamilton's equations of motion\cite{DykmanReview,SchwartzJRSI2011}: 
\begin{align}
\label{eq:HamiltonsEquations}
\dot{x_{i}}=\frac{\partial H}{\partial p_{i}}, \;\; \dot{\phi}_{ij}=\frac{\partial H}{\partial m_{ij}}, \;\;  \dot{p_{i}}=-\frac{\partial H}{\partial x_{i}}, \;\; \dot{m}_{ij}=-\frac{\partial H}{\partial \phi_{ij}}. 
\end{align}

In general, rare events are characterized by ($\vec{x}$, $\vec{p}$, $\boldsymbol{\phi}$, $\boldsymbol{m}$)-trajectories of the LDT equations of motion (e.g., Eq.(\ref{eq:HamiltonsEquations})) with particular boundary conditions, and their probabilities given by Eq.(\ref{eq:QuenchedDistribution}) and Eq.(\ref{eq:Action}) to logarithmic accuracy. Since such solutions {\it minimize the action}, they are most probable, or optimal solutions\cite{DykmanReview}. 

In the case of extinction from a metastable endemic state with constant infection rate, $\beta\!\!=\!\!\beta_{0}$, the solution has a special form, since there is approximately no time-dependence to $\rho(\vec{x},\boldsymbol{\phi};\!\beta)$\cite{FerreiraPRE2016}. Hence $\partial S/\partial t\!=\!H(\vec{x},\boldsymbol{\phi},\vec{p},\boldsymbol{m};\!\beta)\!=\!0$ \cite{DykmanReview,KamenevPRL2008}. Metastability, therefore, implies that network extinction is a zero-energy solution of Eq.(\ref{eq:HamiltonsEquations}) that starts at an endemic state (a fixed point of Eq.(\ref{eq:HamiltonsEquations}) with $\vec{x}\!\neq\!\vec{0}$, $\boldsymbol{\phi}\!\neq\!\boldsymbol{0}$) and ends at the extinct state (a fixed point of Eq.(\ref{eq:HamiltonsEquations}) with $\vec{x}\!=\!\vec{0}$, $\boldsymbol{\phi}\!=\!\boldsymbol{0}$). These boundary conditions are sufficient to determine solutions numerically\cite{LindleyIAMM,FN3}. 

In order to highlight features of the combined WKB and PQT techniques, we look at two example networks where $\beta\!=\!\beta_{0}$, before moving on to the case where external noise is present. The first example is a random network where all nodes have $k$ neighbors ($k$-regular)\cite{MattaEPL2013}. Such networks are useful models of homogeneous topologies with local interactions, and where the fully connected graph can be recovered as $k\!\rightarrow\!N-1$. As written, the dynamical system Eq.(\ref{eq:HamiltonsEquations}) is high dimensional. However, since all nodes have the same degree, we can significantly reduce the dimension by approximating local variables by their averages: $x\!=\!\sum_{i}x_{i}/N$, $\phi\!=\!\sum_{ij}\!\phi_{ij}/[Nk]$, $p\!=\!\sum_{i}p_{i}/N$, and $m\!=\!\sum_{ij}\!m_{ij}/[Nk]$. Assuming that $x_{i}\!=\!x$, $\phi_{ij}\!=\!\phi$, $p_{i}\!=\!p$, $m_{ij}\!=\!m$ for all $i$ and $j$ in the network, the Hamiltonian becomes $H(\vec{x},\boldsymbol{\phi},\vec{p},\boldsymbol{m};\!\beta)=H_{k}(x,\phi,p,m;\!\beta)$ with
\begin{align}
\label{KregHamiltonian}
&H_{k}(x,\phi,p,m;\!\beta)=N\alpha x\big[\!\exp\!\big\{\!\!-\!p\!+\!km(x\!-\!2\phi)/x\!\big\}-1\big]+\nonumber \\
&Nk\beta\phi\big[\!\exp\!\big\{p\!-\!m\!+\!(k\!-\!1)m(1\!-\!x\!-\!2\phi)/(1\!-\!x)\!\big\}-1\big]. 
\end{align}
From Eq.(\ref{eq:HamiltonsEquations}) it follows that $\dot{x}\!\!=\!N^{-1}\partial H/\partial p$, $\dot{p}\!=\!-N^{-1}\partial H/\partial x$, $\dot{\phi}\!=\![Nk]^{-1}\partial H/\partial m$, and $\dot{m}\!=\!-[Nk]^{-1}\partial H/\partial \phi$. Hence, the equations of motion for $\dot{x}$ and $\dot{\phi}$ are:
\begin{align}
\label{KRegEquations}
\dot{x}=&-\alpha x \exp\!\big\{\!\!-\!p\!+\!km(x\!-\!2\phi)/x\!\big\}+\nonumber \\
&\beta k\phi\exp\!\big\{p\!-\!m\!+\!(k\!-\!1)m(1\!-\!x\!-\!2\phi)/(1\!-\!x)\!\big\},\\
\label{KRegEquations2}
\dot{\phi}=&\;\alpha\big[x-2\phi\big]\exp\!\big\{\!\!-\!p\!+\!km(x\!-\!2\phi)/x\!\big\}+\nonumber \\
&\beta\phi\Big[\frac{(k\!-\!1)(1\!-\!x\!-\!2\phi)}{(1\!-\!x)}\!-\!1\Big]\exp\!\big\{p\!-\!m\!\nonumber\\
&+\!(k\!-\!1)m(1\!-\!x\!-\!2\phi)/(1\!-\!x)\!\big\}. 
\end{align}

We point out that the known pair-approximation equations for $k$-regular networks are found by setting the momenta equal to zero in Eqs.(\ref{KRegEquations})-(\ref{KRegEquations2}), $p\!=\!m\!=0$\cite{MattaEPL2013}. In fact, this is a general feature of our approach: {\it network mean-field theories are zero-momentum invariant manifolds of Hamilton's equations of motion}\cite{HindesSR2017}. In general, rare events are describable as trajectories that lie off of this manifold, i.e., $\vec{p}\!\neq\!\vec{0}$ and $\boldsymbol{m}\!\neq\!\boldsymbol{0}$ \cite{SchwartzJRSI2011,FN4}. 

The second example is a star network ($\mathcal{S}(L)$), which is composed of a single central node sharing an edge with $L\!=\!N\!-\!1$ ``leaf" nodes. Leaf nodes do not share edges. Star networks are relevant because they exhibit localization of the endemic state around the central node near threshold. Localization is ubiquitous and is observed, for instance, in certain classes of random networks with unbounded degree distributions as $N\!\rightarrow\!\infty$\cite{GoltsevPRL2012,Mata2015PRE,FerreiraPRE2016,FerreiraPRE2016_2}. 

Let us denote the infection and momentum of the central node with index $0$, $x_{0}$ and $p_{0}$, respectively. Similar to the $k$-regular networks, we define average coordinates for the leaves: $x_{L}\!=\!\sum_{j\in\mathcal{L}}x_{j}/L$, $\phi_{L0}\!=\!\sum_{j\in\mathcal{L}}\phi_{j0}/L$, $\phi_{0L}\!=\!\sum_{j\in\mathcal{L}}\phi_{0j}/L$, $p_{L}\!=\!\sum_{j\in\mathcal{L}}p_{j}/L$, $m_{L0}\!=\!\sum_{j\in\mathcal{L}}m_{j0}/L$, and $m_{0L}\!=\!\sum_{j\in\mathcal{L}}m_{0j}/L$, where $\mathcal{L}$ denotes the set of leaves. Since there are two types of nodes and only one type of edge in a $\mathcal{S}(L)$-- between the central node and each leaf-- we can apply the constraint $\!\phi_{0L}\!=\!x_{L}-x_{0}+\phi_{L0}$ with $m_{0L}\!=\!0$\cite{FN4}. Approximating all leaf variables by their respective averages gives the $\mathcal{S}(L)$ Hamiltonian 
\begin{align}
\label{Star Hamiltonian}
&H_{\text{star}}(\vec{x},\boldsymbol{\phi},\vec{p},\boldsymbol{m};\!\beta)=\alpha  x_{0}\!\big[\!\exp\{\!-p_{0}\!-\!L\phi_{L0}m_{L0}/x_{0}\!\}\!-\!1\big]\nonumber\\
&+\alpha Lx_{L}\!\big[\!\exp\{\!-p_{L}\!+\!(x_{0}\!-\!\phi_{L0})m_{L0}/x_{L}\!\}\!-\!1\big]+\nonumber \\
&\beta L(x_{L}\!-\!x_{0}\!+\!\phi_{L0})\!\big[\!\exp\{p_{0}\!+\!(L\!-\!1)(1\!-\!x_{L}-\!\phi_{L0})m_{L0}\nonumber\\
&/(1\!-\!x_{0})\}\!-\!1\big]+\beta L\phi_{L0}\!\big[\!\exp\{p_{L}\!-\!m_{L0}\}\!-\!1\big].
\end{align}
\begin{figure}
\includegraphics[scale=0.235]{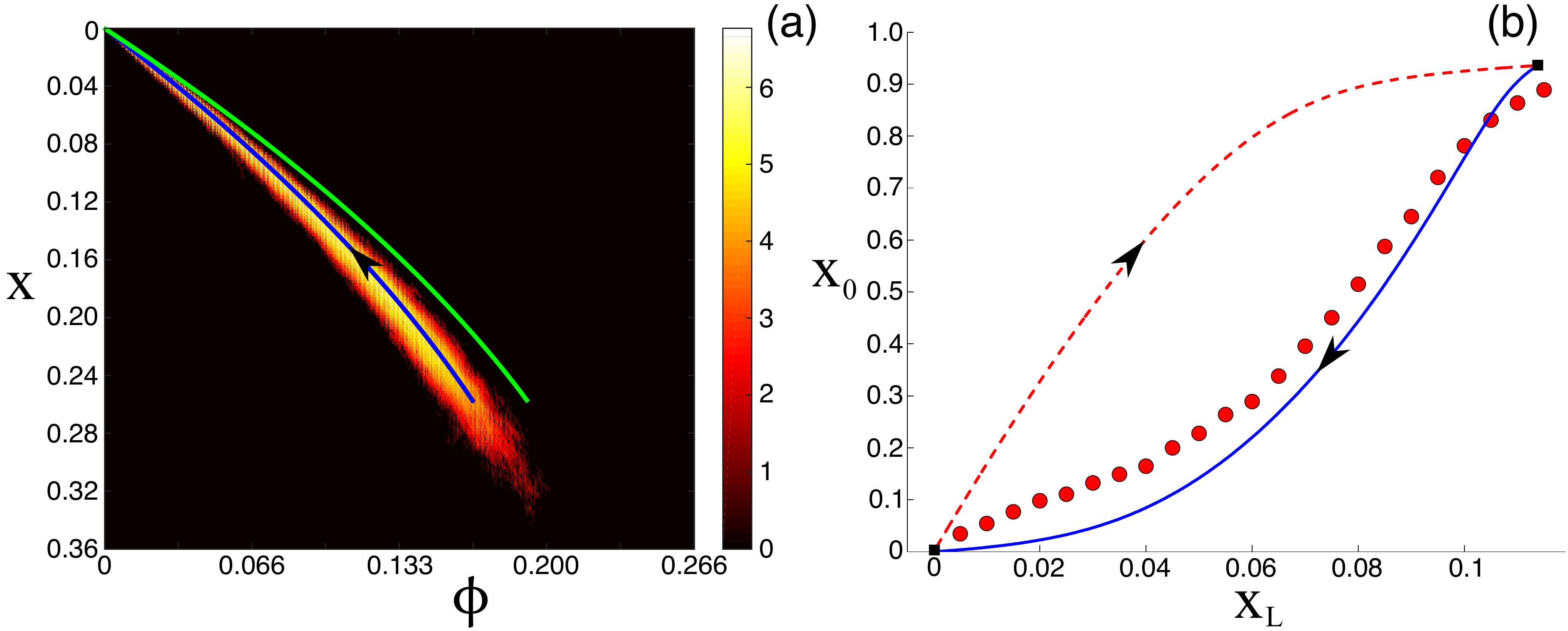}
\caption{Extinction paths in networks with internal noise, $D\!=\!0$. (a) path density on log scale shown as a function of the fraction of infected nodes, $x$, and the fraction of edges connecting infected and susceptible neighbors, $\phi$, in a 6-regular network with $500$ nodes and $\beta_{0}\!=\!0.2580$. The predicted path (blue) is compared to a path assuming independence of neighbors (green). (b) fraction of $10^{4}$ simulated extinction paths where the central node of a star network is infected, $x_{0}$, and a fraction, $x_{L}$, of leaf nodes are infected (red circles) in a network with $1001$ nodes and $\beta_{0}\!=\!0.1373$. The predicted path (blue) is compared to a path that leads to the endemic state (dashed-red). Arrows indicate the direction in time.}
\label{HeatMap}
\end{figure}  
 
Predictions of $\rho(\vec{x},\boldsymbol{\phi},t;\!\beta)$ from solving Eq.(\ref{eq:HamiltonsEquations}) for both networks  agree well with measured histograms, Fig.\ref{Distr} (blue). In addition, Figure \ref{HeatMap} shows path-projections of many Monte-Carlo simulations that precede extinction for both networks\cite{FN3}. Panel (a) is a heat map for a 6-regular network, giving the extinction-path density as a function of the fraction of infected nodes and edges. The predicted path is in blue, which lies on the maximum of the density, as expected. The dashed-green line represents a path that assumes statistical independence of neighbors, $\phi\!=\!x(1-x)$, which is consistent with standard heterogeneous and quenched mean-field approximations\cite{HindesPRE2017,HindesSR2017}; such a path does not lie on the maximum density. Therefore, we see that the PQT provides a significant quantitative improvement over standard mean-field approximations in predicting extinction\cite{MattaEPL2013}. %The path shapes give insight into the dynamical sequences leading to extinction[]

Panel (b) shows the fraction of $10^4$ stochastic simulations in which the central node of a $\mathcal{S}(L)$ is infected while a fraction $x_{L}$ leaf nodes are infected, just before extinction occurs (red circles). The predicted path from Eq.(\ref{eq:HamiltonsEquations}) and Eq.(\ref{Star Hamiltonian}) is shown in blue, and agrees qualitatively well with simulation results. The red-dashed curve is the predicted path into the endemic state ($p_{0}\!=\!p_{L}\!=\!m_{L0}\!=\!0$) from an initial seed infection, $x_{L}\!=\!\epsilon\!\ll\!1$. The dissimilarity between the two paths (only overlapping at the endpoints) demonstrates the highly non-reversible nature of rare events in heterogeneous networks driven by internal noise\cite{HindesPRL2016,HindesPRE2017}. 

\begin{figure}[h]
\includegraphics[scale=0.236]{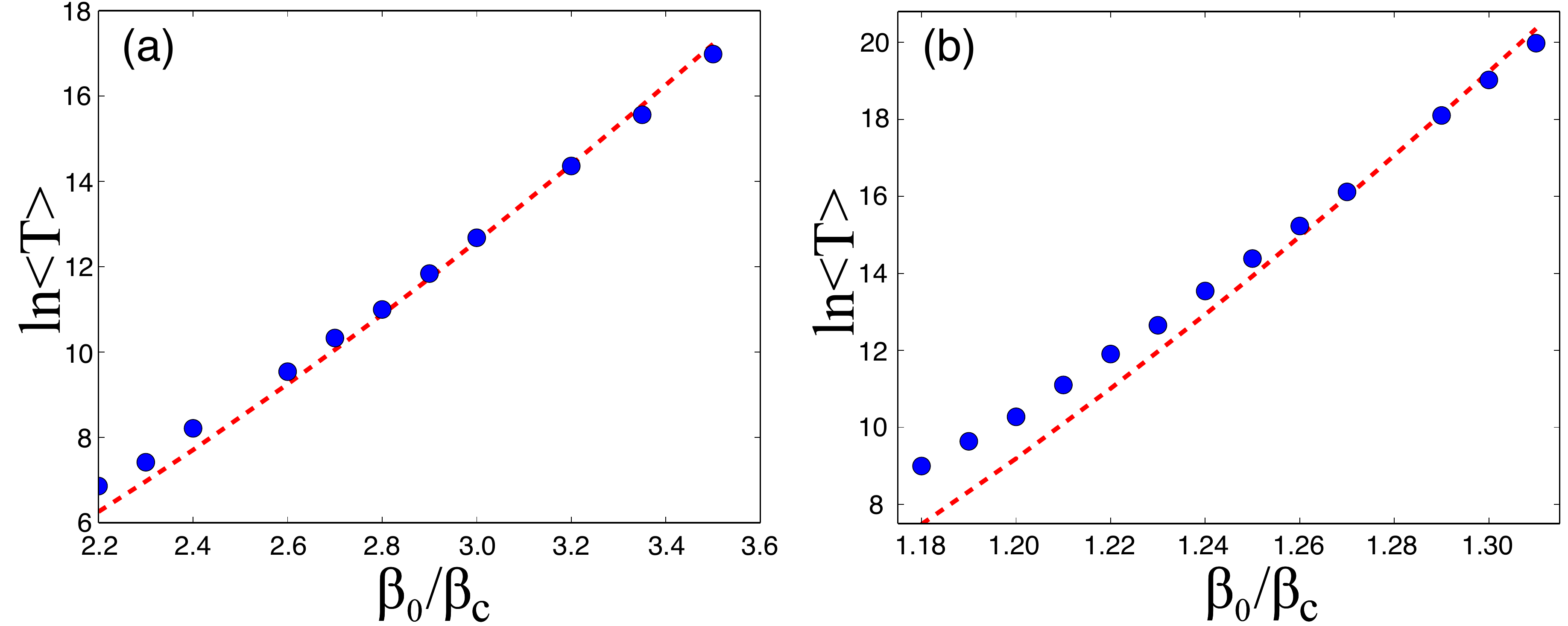}
\caption{Extinction times in networks with internal noise, $D\!=\!0$. Log of the average extinction time (blue circles) versus $\beta_{0}$ (measured in units of the critical value). (a) star network with $1001$ nodes. (b) 6-regular network with $500$ nodes. Predictions are shown with a dashed-red line from Eq.(\ref{eq:Action}) and Eq.(\ref{eq:Time}). The pre-factor in Eq.(\ref{eq:Time}) was fitted separately for both networks.}
\label{Times1}
\end{figure}

In addition to the paths and probabilities of rare events, we are also interested in their frequency (rate). Generally, rare events are Poisson processes with rates proportional to their probabilities\cite{AssafRev,AssafPRE2010,KamenevPRL2008}. Hence the inverse rate of extinction, or average time $\left<T\right>$, is expected to be inversely proportional to the probability of extinction, or to logarithmic accuracy    
\begin{align}
\label{eq:Time}
\ln\!\left<T\right>=S(\vec{x}\!=\!\vec{0},\boldsymbol{\phi}\!=\!\boldsymbol{0};\!\beta)+B, 
\end{align}
where $B$ is assumed to be an $\mathcal{O}(1)$ pre-factor\cite{AssafPRE2010,SchwartzPRE2015}. Figure \ref{Times1} shows the average extinction time as a function of $\beta/\beta_{cr}$\cite{MattaEPL2013} for a 6-regular network and a $\mathcal{S}(L)$ with $D\!=\!0$. Predictions from Eq.(\ref{eq:Time}) are shown in red with good agreement.

So, far we have considered only the effects of internal noise on the occurrence of rare extinction. If $\beta(t)$ has a non-zero noise amplitude, then we must integrate over realizations of $\beta(t)$ to get the marginal distribution of rare network states, $\rho(\vec{x},\boldsymbol{\phi},t)$. In the case where $\beta(t)\!=\!\beta_{0}[1-\xi(t)]$, and $\xi(t)$ is a GWNP, the statistical weight for a given realization of $\xi(t)$ is $\mathcal{P}[\xi(t)]\!\sim\!\exp\{\!-\!\int[\xi^{2}\!/4D]dt\}$\cite{KamenevPRL2008}. Therefore, the marginal distribution takes the functional form:
\begin{align}
\label{eq:Marginal}
&\rho(\vec{x},\boldsymbol{\phi},t)=\!\!\int\!\!\mathcal{D}\xi\;\mathcal{P}[\xi(t)]\rho(\vec{x},\boldsymbol{\phi},t;\!\beta(\xi))\\
&\!=\!\!\!\int\!\!\mathcal{D}\xi\exp\!\Big\{\!\!\!-\!\!\!\int\!\!\!\big[\vec{p}\!\cdot\!\dot{\vec{x}}\;+\!<\!\!\boldsymbol{m},\!\dot{\boldsymbol{\phi}}\!\!>\!-H(\vec{x},\boldsymbol{\phi},\vec{p},\boldsymbol{m};\!\beta)\!+\!\frac{\xi^{2}}{4D}\big]dt\!\Big\}\nonumber.  
\end{align}

As we have seen, rare events occur with exponentially small probabilities along optimal paths described by analytical mechanics. Leveraging this insight once again, we approximate Eq.(\ref{eq:Marginal}) by its maximum-probability contribution. We note that the exponent in Eq.(\ref{eq:Marginal}) takes the form of an effective action, $S_{\text{eff}}(\vec{x},\boldsymbol{\phi},\xi,t)$, whose effective Hamiltonian is 
\begin{align}
\label{EffectiveHamiltonian}
H_{\text{eff}}(\vec{x},\boldsymbol{\phi},\vec{p},\boldsymbol{m},\xi)=H(\vec{x},\boldsymbol{\phi},\vec{p},\boldsymbol{m};\!\beta(\xi))-\frac{\xi^2}{4D}. 
 \end{align}
Therefore, the minimum action solution satisfies another Hamilton-Jacobi equation, $\partial S_{\text{eff}}/\partial t +H_{\text{eff}}\!=\!0$, with $H_{\text{eff}}\!=\!0$. %(since the endemic state is metastable)

When external noise is a GWNP and $H_{\text{eff}}(\vec{x},\boldsymbol{\phi},\vec{p},\boldsymbol{m},\xi)$ is quadratic in $\xi$, we can solve explicitly for the realization of $\xi$ with maximum probability as a function of the optimal path coordinates, $\xi(t)\!=\!\bar{\xi}(\vec{x},\boldsymbol{\phi},\vec{p},\boldsymbol{m})$. Substituting, $\beta\!=\bar{\beta}\!=\!\beta_{0}[1-\bar{\xi}]$ into Eq.(\ref{eq:Hamiltonian}) and Eq.(\ref{EffectiveHamiltonian}), with the latter set equal to zero, we find\!\cite{FN1}:
\begin{align}
\label{OptimalNoise}
&\bar{\xi}=-2D\beta_{0}\!\sum_{ij}\!A_{ij}\phi_{ij}\!\bigg(\!\!\exp\!\Big\{p_{i}\!-\!m_{ij} +\\
&\!\sum_{l}\!A_{il}\frac{(1\!-\!\delta_{jl})}{(1\!-\!x_{i})}\!\big[\!-\!\phi_{il}m_{il}+(1\!-\!x_{i}\!-\!\phi_{il})m_{li}\big]\!\Big\}\!-\!1\!\bigg).\nonumber  
\end{align}

As with the $D\!=\!0$ case, optimal paths with external noise satisfy Eq.(\ref{eq:HamiltonsEquations}). The main difference is that the infection rate implicit in Eq.(\ref{eq:HamiltonsEquations}) is set equal to the optimal realization, $\beta(t)\!=\!\bar{\beta}$ \cite{FN5}. The fixed-point boundary conditions unchanged. Just as before, the action is computed along the path,    
\begin{align}
\label{FinalAction}
S_{\text{eff}}\big(\vec{x},\boldsymbol{\phi};\bar{\beta}\!\;\big)=\int\!\!\big[\vec{p}\cdot\dot{\vec{x}}\;+\!<\!\!\boldsymbol{m},\!\dot{\boldsymbol{\phi}}\!\!>\!\big]dt,  
\end{align}
giving average extinction times:
\begin{align}
\label{eq:Time2}
\ln\!\left<T\right>=S_{\text{eff}}(\vec{x}\!=\!\vec{0},\boldsymbol{\phi}\!=\!\boldsymbol{0};\!\bar{\beta})+B.      
\end{align}  

\begin{figure}[t]
\includegraphics[scale=0.235]{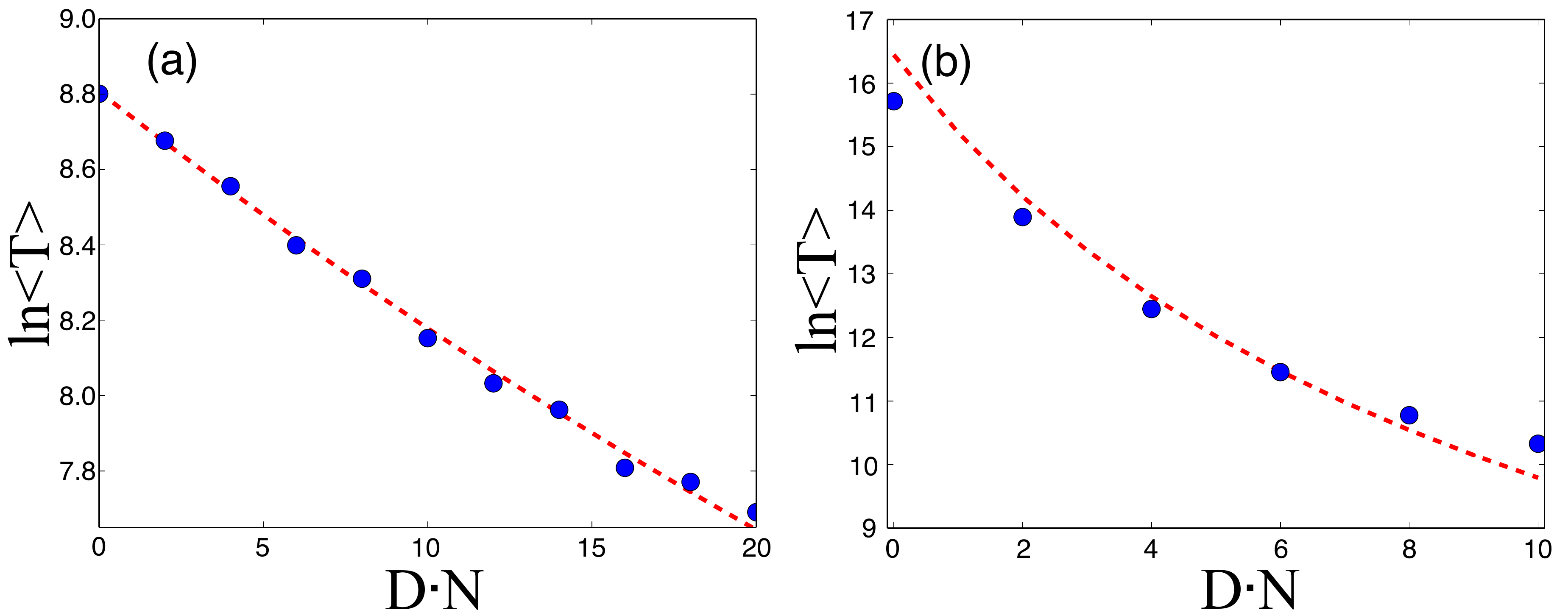}
\caption{Extinction times in networks with internal and external noise. Log of the average extinction time (blue circles) versus external noise amplitude $D$ (measured in units of $1/N$)\cite{FN2}. (a) star network with $1001$ nodes and $\beta_{0}\!=\!0.1144$. (b) 8-regular network with $500$ nodes and $\beta_{0}\!=\!0.1786$. Predictions are shown with a dashed-red line from Eq.(\ref{FinalAction}) and Eq.(\ref{eq:Time}). The pre-factor in Eq.(\ref{eq:Time}) was fitted separately for both networks.}
\label{Times2}
\end{figure}

\begin{figure}[h]
\includegraphics[scale=0.2369]{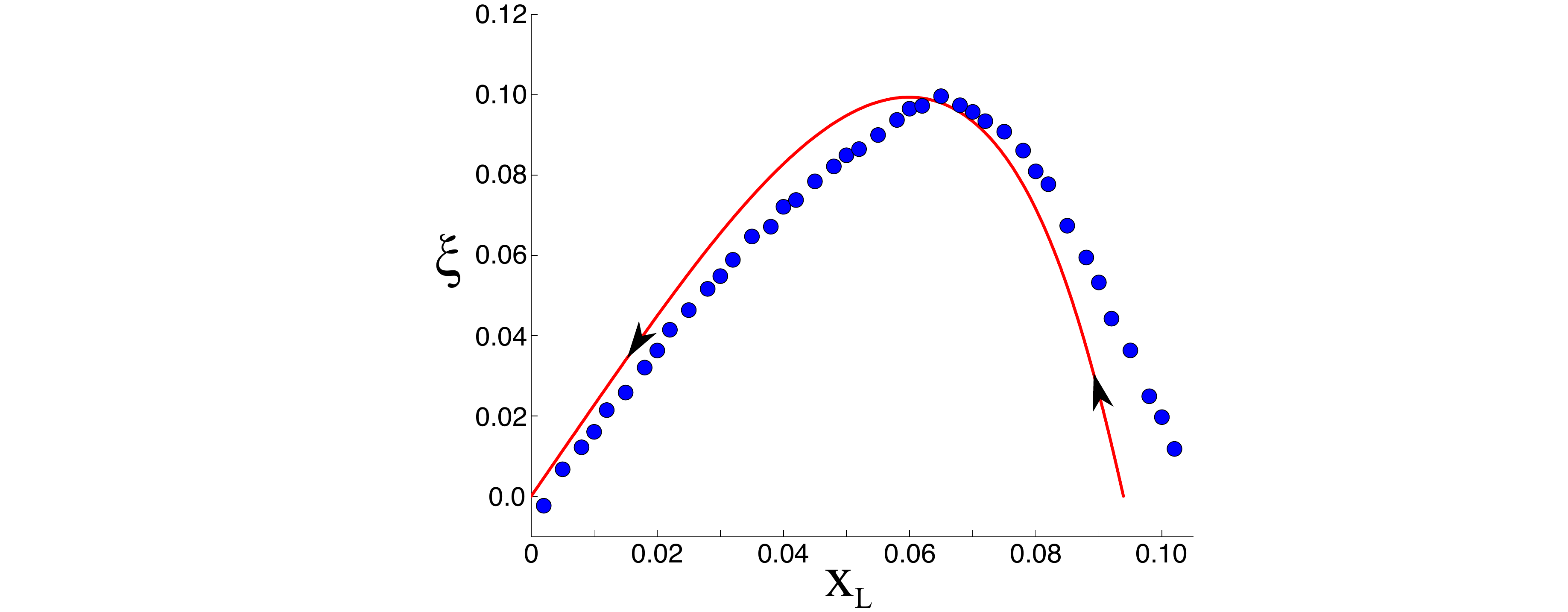}
\caption{Optimal realization of external noise leading to extinction. Average value of $\xi$ when a fraction, $x_{L}$, of leaf nodes are infected in a star network (blue circles) with $1001$ nodes and $\beta_{0}\!=\!0.1144$. The average for each point was computed from $5\!*\!10^{4}$ extinction paths. The predicted path is shown in red from Eq.(\ref{OptimalNoise}). Arrows indicate the direction in time.}
\label{ONR}
\end{figure}

Predictions of $\ln\!\left<T\right>$ from Eq.(\ref{eq:Time2}) are compared to simulations in Fig.\ref{Times2}, with changing $D$ and fixed $\beta_{0}$, and show good agreement for the two example networks\cite{FN3}. Qualitatively, we find that $\ln\!\left<T\right>$ decreases as $D$ increases -- demonstrating that large fluctuations producing extinctions in networks are only enhanced by external noise. In addition, the shape of the optimal external noise in simulations can be compared directly to predictions, Eq.(\ref{OptimalNoise}). For example, Fig.\ref{ONR} shows the generic form of $\xi$, in which the external noise increases from zero at the endemic state, reaches a maximum at some intermediate infection level, and then decreases to zero at the extinct state.

\begin{figure}[h]
\includegraphics[scale=0.236]{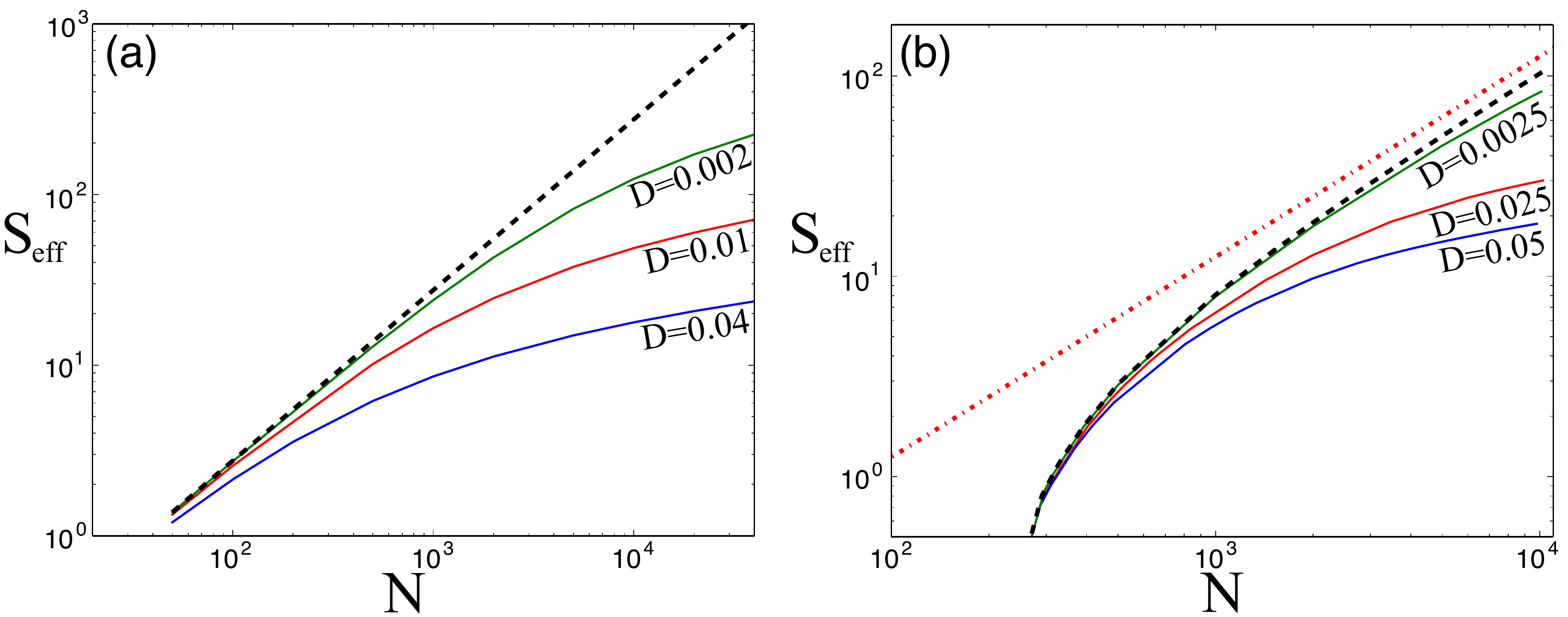}
\caption{Cross-over in $S_{\text{eff}}$ as a function of network size for different external noise intensities. (a) 8-regular network with $\beta_{0}\!=\!0.1786$. (b) star network with $\beta_{0}\!=\!0.1144$. Black-dashed lines correspond to $D\!=\!0$ in (a) and (b). Linear scaling is shown with a red-dashed line in (b).} %%%%%%%%%%%%%%%%%%%%%\textcolor{red}{More points for b ?}}
%%%%%%%%%%%%%%%%%%%%%
%%%%%%%%%%%%%%%%%%%%%
%%%%%%%%%%%%%%%%%%%%%
%%%%%%%%%%%%%%%%%%%%%
%%%%%%%%%%%%%%%%%%%%%
%%%%%%%%%%%%%%%%%%%%%
%%%%%%%%%%%%%%%%%%%%%
%%%%%%%%%%%%%%%%%%%%%
%%%%%%%%%%%%%%%%%%%%%
\label{CrossOver}
\end{figure}

Given the outlined mechanics for rare dynamics in networks with internal and external noise, we may wonder how each differs in its effect on extinction. For instance, it is well known that the probability of large deviations from internal noise generally decreases exponentially with the network size, $N$, such that $S\!\sim\!N$ as $N\!\!\rightarrow\!\!\infty$ \cite{DykmanReview,AssafRev,HindesPRE2017}. An interesting question concerns how $S_{\text{eff}}(\vec{x}\!=\!\vec{0},\boldsymbol{\phi}\!=\!\boldsymbol{0};\!\bar{\beta})$ (or $S_{\text{eff}}$ for short) scales with $N$ in the presence of external noise and different network topology. Figure \ref{CrossOver} shows $S_{\text{eff}}$ as a function of $N$ for two example networks and several values of $D$. Panel (a) shows results for an $8$-regular network. For $D\!\neq\!0$, we can see that $S_{\text{eff}}$ scales linearly with $N$ (black-dashed line) for small $N$, implying that for sufficiently small networks, internal noise is dominant and external noise has a only a perturbative effect. Therefore for k-regular networks there is always some range of $N$ over which $S_{\text{eff}}$ scales linearly with $N$ for any $D$. As $N$ increases, however, their is a {\it crossover to $S_{\text{eff}}\!\sim\!1/D$}, as the intensity of internal fluctuations becomes negligible and extinction depends effectively on fluctuations in $\beta$ alone\cite{LoraPRL2010}.
%with continuously varying, sub-linear scaling, $S_{\text{eff}}\sim N^{\alpha(\beta,D)}$ with $\alpha\!\leq\!1$}, as the intensity of internal fluctuations become small and extinction depends effectively on $D$ alone. As $N\!\!\rightarrow\!\!\infty$, $S_{\text{eff}}\!\sim\!1/D$. 

On the other hand, panel (b) shows the result for a $\mathcal{S}(L)$. In contrast, there is a minimum network size for fixed $\beta_{0}$, $N_{c}(\beta_{0})$, such that smaller stars cannot stabilize endemic states. As a consequence, {\it $S_{\text{eff}}$ grows nonlinearly from zero for $N\!\gtrsim\!N_{c}$} (black-dashed line), even when $D\!=\!0$. The linear scaling (red-dashed line) is found when $D\!=\!0$, but only asymptotically as $N\!\!\rightarrow\!\!\infty$, as the dashed lines approach each other. The internal noise scaling pattern for a $\mathcal{S}(L)$ is a consequence of localization near threshold, and we expect it to appear in networks that show localization more generally \cite{FN4}. For $D\!\neq\!0$, there is a cross-over from the internal noise scaling pattern to $S_{\text{eff}}\!\sim\!1/D$ as $N\!\!\rightarrow\!\!\infty$, just as in the $k-$regular case. However, since the linear scaling with $N$ is reached only asymptotically in the absence of external noise, it is possible that such scaling is never found if the external noise is too large, e.g., $D\!=\!0.05$ in Fig.\ref{CrossOver}(b).

We conjecture that the noise cross-over for localized endemic states should be observed in large power-law networks with degree exponents greater than three. This is because the endemic state near threshold\cite{FN6} in such networks is known to behave like a star sub-graph of the full network with $L\!\rightarrow\!k_{max}$, where $k_{max}$ is the maximum degree of the hub node\cite{MattaEPL2013}.   

In conclusion, we have developed an analytical framework that describes the most-probable pathway of rare events, such as extinction, in networks having both internal and external noise. Our formalism combined pair-quenched mean-field approximations for network dynamics with WKB techniques and large deviation theory. Using these techniques, we were able to predict the distribution of rare network states and the frequency of extinction, depending on both a network's topology and the intensity of external noise. We showed that the presence of any external noise leads to a cross-over region where the familiar exponential dependence of large deviations on a network's size ($N$) is lost -- in the infinite-size limit, depending only on the external noise intensity, $D$. 

We observed that the cross-over occurs as a network's probability exponent approaches a scale of $1/D$. As a consequence, there may be no range of $N$ over which the exponential scaling with $N$ is found for topologies that are localized near threshold, since such networks show the exponential scaling with $N$ only asymptotically as $N\!\rightarrow\!\infty$ (without external noise). With external noise, the cross-over may occur for $N$ that is too small to show exponential scaling.    
%%%%%%%
%%%%%%%
%%%%%%%
%%%%%%%   
%%%%%%%\textcolor{red}{[Supplementary material: near threshold calculations for both networks? Derivation of HJE?]}\\
%%%%%%%
%%%%%%%
%%%%%%%
\acknowledgments
\noindent JH is a National Research Council postdoctoral fellow. IBS was supported by the U.S. Naval Research Laboratory funding (N0001414WX00023) and office of Naval Research (N0001416WX00657) and (N0001416WX01643). We thank L. Mier-Y-Teran-Romero for useful discussions.

\end{document}